\documentclass[conference]{IEEEtran}
\IEEEoverridecommandlockouts
\usepackage{cite}
\usepackage{amsmath,amssymb,amsfonts}
\usepackage{algorithmic}
\usepackage{graphicx}
\usepackage{textcomp}
\usepackage{xcolor}
\def\BibTeX{{\rm B\kern-.05em{\sc i\kern-.025em b}\kern-.08em
    T\kern-.1667em\lower.7ex\hbox{E}\kern-.125emX}}
\begin{document}

\title{Equivalent DQ Sequence-Domain Model of Unbalanced Three-Phase Passive Elements for Power Converter Controllers\\
\thanks{\copyright{} 2023 IEEE. Personal use of this material is permitted.
Permission from IEEE must be obtained for all other uses, in any current or
future media, including reprinting/republishing this material for advertising
or promotional purposes, creating new collective works, for resale or
redistribution to servers or lists, or reuse of any copyrighted component of
this work in other works.\\
DOI: 10.1109/APEC43580.2023.10131365}
}

\author{\IEEEauthorblockN{Airán Francés}
\IEEEauthorblockA{\textit{Centro de Electrónica Industrial} \\
\textit{Universidad Politécnica de Madrid}\\
Madrid, Spain \\
airan.frances@upm.es}
\and
\IEEEauthorblockN{Luis Saz}
\IEEEauthorblockA{\textit{Centro de Electrónica Industrial} \\
\textit{Universidad Politécnica de Madrid}\\
Madrid, Spain \\
la.saz@alumnos.upm.es}
\and
\IEEEauthorblockN{Rafael Castillo}
\IEEEauthorblockA{\textit{WEMPEC Research Group} \\
\textit{University of Wisconsin}\\
Madison, United States \\
castillosier@wisc.edu}
\and
\IEEEauthorblockN{Dionisio Ramirez}
\IEEEauthorblockA{\textit{Centro de Electrónica Industrial} \\
\textit{Universidad Politécnica de Madrid}\\
Madrid, Spain \\
dionisio.ramirez@upm.es}
\and
\IEEEauthorblockN{Javier Uceda}
\IEEEauthorblockA{\textit{Centro de Electrónica Industrial} \\
\textit{Universidad Politécnica de Madrid}\\
Madrid, Spain \\
javier.uceda@upm.es}
}

\maketitle

\begin{abstract}
Equivalent $dq0$ models of three-phase passive impedances are widely used for the design of three-phase inverter and rectifier controllers in the synchronous frame. However, these equivalent models assume that the impedance in each phase are the same, which is not always applicable. The fact that the impedance in each phase is different, generates second order harmonic content in the $dq0$ components, which hinders the advantages of the $dq0$ transformation. To avoid the second harmonic in these scenarios, it is possible to use the Fortescue’s theorem to represent any asymmetrical set of three-phase signals as a linear combination of three symmetrical sequences. In the literature, only a few works have attempted to derive the equivalent model of unbalanced three-phase passive elements in the sequence domain, and none is able to account for the interaction between the zero sequence and the positive and negative sequences. This work presents the derivation of the equivalent $dq$ sequence-domain model of unbalanced three-phase passive elements both analytically and with an intuitive graphical approach. Experimental results are shown to validate the proposed models. 
\end{abstract}


\section{Introduction}
Many works related to the design of controllers for three-phase inverters and rectifiers can be found in the literature. These controllers can be implemented in the alpha-beta plane \cite{Control_stationary,grid_voltage_estimation_stationary,resonant_control_stationary}, the $dq$ synchronous frame \cite{Input_voltage_distortion_dq,virtual_time_constant_dq,decentralized_control_dq}, or the sequence-domain framework \cite{impedance_modeling_sequence,Model_predictive_sequence,Dual_current_control_sequence}. Most of them consider the passive impedances to be balanced in the analysis.

However, in some cases, it is not true due to tolerances in components, especially magnetics, and in others, it is not true by design. In those cases, the errors in the model will become perturbations in the feedback loop, which will affect the regulator's performance.
Only a few works have attempted to derive a $dq$ sequence-domain equivalent model of three-phase unbalanced passive impedances \cite{unbalanced_impedance_original,adaptive_control_unbalanced_impedance}, however, they have not taken into account the zero sequence properly. According to that models, a set of voltages without zero sequence applied to an unbalanced impedance cannot generate current with zero sequence and vice versa, which is not correct. 

This work derives the $dq$ sequence-domain equivalent model of any unbalanced impedance considering the interaction between positive, negative, and zero sequences. 

\section{Equivalent model inference}
The sequence model is based on the Fortescue’s theorem which states that an asymmetrical three-phase signal can be broken down into three symmetrical sequences, namely the positive ($v_{abc}^+$), negative ($v_{abc}^-$), and zero ($v_{abc}^0$) sequences.
Furthermore, these sequences can be derived from the asymmetrical time-domain signals using the Lyon transformation \cite{Teodorescu_Book_2010}:

\begin{equation} \label{Lyon}
    \left[ \begin{array}{c}
v_a^+ \\
v_a^- \\
v_a^0 
\end{array} \right]=[T]v_{abc}= \dfrac{1}{3}
\left[ \begin{array}{ccc}
1 & a & a^2 \\
1 & a^2 & a \\
1 & 1 & 1
\end{array} \right]
\left[ \begin{array}{c}
v_a \\
v_b \\
v_c 
\end{array} \right]
\end{equation}

\noindent where $a$ represents a 120º phase-shift. It is then possible to utilize the Clarke, $[T_c]$ and Park transformations, $[T_p(\omega)]$ on each of the sequences (positive, negative, and zero) to obtain their equivalent sequences in the $dq$ rotating reference frame. 

\begin{equation} \label{Park1}
   v_{dq} = \left[ \begin{array}{c}
v_d \\
v_q 
\end{array} \right]=[T_p(\omega)] \cdot [T_c] \cdot v_{abc}
\end{equation}

\noindent where

\begin{equation}
    [T_c] = \dfrac{2}{3} \left[ \begin{array}{ccc}
    1 & -\dfrac{1}{2} & -\dfrac{1}{2} \\ \\
    0 & \dfrac{\sqrt{3}}{2} & -\dfrac{\sqrt{3}}{2}
    \end{array} \right]
\end{equation}

\begin{equation} \label{Park2} 
[T_p(\omega)]= \left[ \begin{array}{cc}
cos(\omega t) &  sin(\omega t)\\
-sin(\omega t) & cos(\omega t)
\end{array} \right]
\end{equation}

An important advantage of the $dq$ components of these sequences is that, even if the signal is asymmetrical and unbalanced, they are constant in the steady state. In the following sections, the derivation of a sequence-domain equivalent circuit for any given impedance is detailed.

\subsection{Derivation of the Resistive Equivalent Impedance}

A purely resistive impedance is the least complex scenario and is therefore a good starting point for the construction of the equivalent impedance model. The three-phase equation of a resistive impedance can be expressed by the following expression:

\begin{equation} \label{Rabc}
    \left[ \begin{array}{c}
v_a \\
v_b \\
v_c 
\end{array} \right]= 
\left[ \begin{array}{ccc}
R_a & 0 & 0 \\
0 & R_b & 0 \\
0 & 0 & R_c
\end{array} \right]
\left[ \begin{array}{c}
i_a \\
i_b \\
i_c 
\end{array} \right] = \left[ R_{abc} \right] \cdot i_{abc}
\end{equation}

In the case of a balanced impedance, the matrix $[R_{abc}]$ will be diagonal and $R_a=R_b=R_c=R$. In that case, it is possible to operate using the Clarke transform, $\left[ T_{c} \right]$, to obtain the equivalent impedance matrix, $\left[ R_{\alpha \beta} \right]$, in the $\alpha\beta$ reference frame:

\begin{equation} \label{R_en_ab}
     v_{\alpha \beta} = \left[ T_{c} \right] \left[ R_{abc}  \right] \left[ T^{-1}_{c} \right] \cdot i_{\alpha \beta} = \left[ R_{\alpha \beta} \right] \cdot i_{\alpha \beta}
\end{equation}

The $\alpha \beta$ components can be expressed in the $dq$ framework by means of a rotation of value $\theta= \omega t$, which can be represented as: 

\begin{equation} \label{ab_en_dq}
    v_{\alpha \beta} =v_{dq} \cdot e^{j\omega t}
\end{equation}

Substituting (\ref{ab_en_dq}) in (\ref{R_en_ab}), the impedance equivalent matrix, $R_{dq}$ can be expressed in the $dq$ framework:

\begin{equation}
    v_{dq} \cdot e^{j\omega t} = \left[ R_{dq} \right] \cdot i_{dq} \cdot e^{j\omega t}
\end{equation}

Then, the phasor equation of the equivalent circuit in the $dq$ reference frame is:

\begin{equation}
     v_{dq} = \left[ R_{dq} \right] \cdot i_{dq}
\end{equation}

\noindent being 

\begin{equation}
    \left[ R_{dq} \right] =  \left[ R_{\alpha \beta} \right] = R \cdot I
\end{equation}

\noindent where I is the identity matrix. 

The case of an unbalanced impedance is of higher complexity since in that case, it will be necessary to utilize the positive, negative, and zero sequences of the three-phase system. The goal is to find a 6x6 equivalent impedance matrix $[R^{+-0}_{dq}]$  to model the interaction between voltage and current across all sequences of the equivalent circuit.

\begin{equation} \label{R_equivalent}
    \left[ \begin{array}{c}
v_d^+ \\
v_q^+ \\
v_d^- \\
v_q^- \\
v_d^0 \\
v_q^0 
\end{array} \right]=v_{dq}^{+-0}=[R^{+-0}_{dq}]i_{dq}^{+-0} = [R^{+-0}_{dq}]
\left[ \begin{array}{c}
i_d^+ \\
i_q^+ \\
i_d^- \\
i_q^- \\
i_d^0 \\
i_q^0 
\end{array} \right]
\end{equation}

In this case, the Lyon transformation can be used to express the unbalanced system as the sum of the three balanced sequences. First, the impedance expression is extended to include the different sequences as follows:

\begin{equation}
        v^{+-0}_{abc} = \left[ \begin{array}{c}
        v^+_{abc} \\
        v^-_{abc} \\
        v^0_{abc}
        \end{array} \right]= [R^{+-0}_{abc}] \cdot \left[ \begin{array}{c}
        i^+_{abc} \\
        i^-_{abc} \\
        i^0_{abc} 
        \end{array} \right] = [R^{+-0}_{abc}] \cdot i^{+-0}_{abc}
\end{equation}

\noindent where $v^+$, $v^-$, and $v^0$ are the vectors of the $abc$ signals of the positive, negative, and zero sequences, respectively, and $[R^{+-0}_{abc}]$ is the extended impedance, which takes into account how the different voltages sequences are affected by each current sequence. Notice that, in this analysis, the zero sequence is also represented with $d$ and $q$ components, even though it is a common mode signal. The approach is equivalent to a single-phase signal, it is possible to create an artificial signal in quadrature, using a Quadrature Signal Generator (QSG), and use it as the beta component to create the $dq$ components. The reason to do that is to be able to account for the phase of the zero sequence, compared with the positive sequence.

Then the Clark transformation is applied, resulting in:

\begin{equation}
    v^{+-0}_{\alpha \beta} = [R^{+-0}_{\alpha \beta}] \cdot i^{+-0}_{\alpha \beta}
\end{equation}

Finally, using the Park transformation, the following expression is obtained:

\begin{equation}
    v_{dq}^{+-0} \cdot e^{j\omega t} =[R^{+-0}_{dq}]i_{dq}^{+-0} \cdot e^{j\omega t}
\end{equation}

The phasor equation of the equivalent circuit in the sequence domain is (\ref{R_equivalent}).

In the following, a graphical interpretation of these steps is detailed. The method used to construct the matrix $[R^{+-0}_{dq}]$ is based on the superposition principle. A unit-value $d$ and $q$ component of each sequence of the current is applied to calculate its contribution to each sequence of the voltages. These unit currents were calculated as symmetrical three-phase currents for the positive and negative sequences, and as a common mode current for the zero sequence.  

These unit-value currents were used to calculate the resulting voltages when applied to a generic unbalanced resistive impedance. The relationships among each of the currents injected by the superposition principle and the corresponding voltages involved are used to fill in each of the columns in matrix $[R^{+-0}_{dq}]$.

\begin{figure}[htbp] \label{phasores_R_Id}
\centerline{\includegraphics[width=88mm]{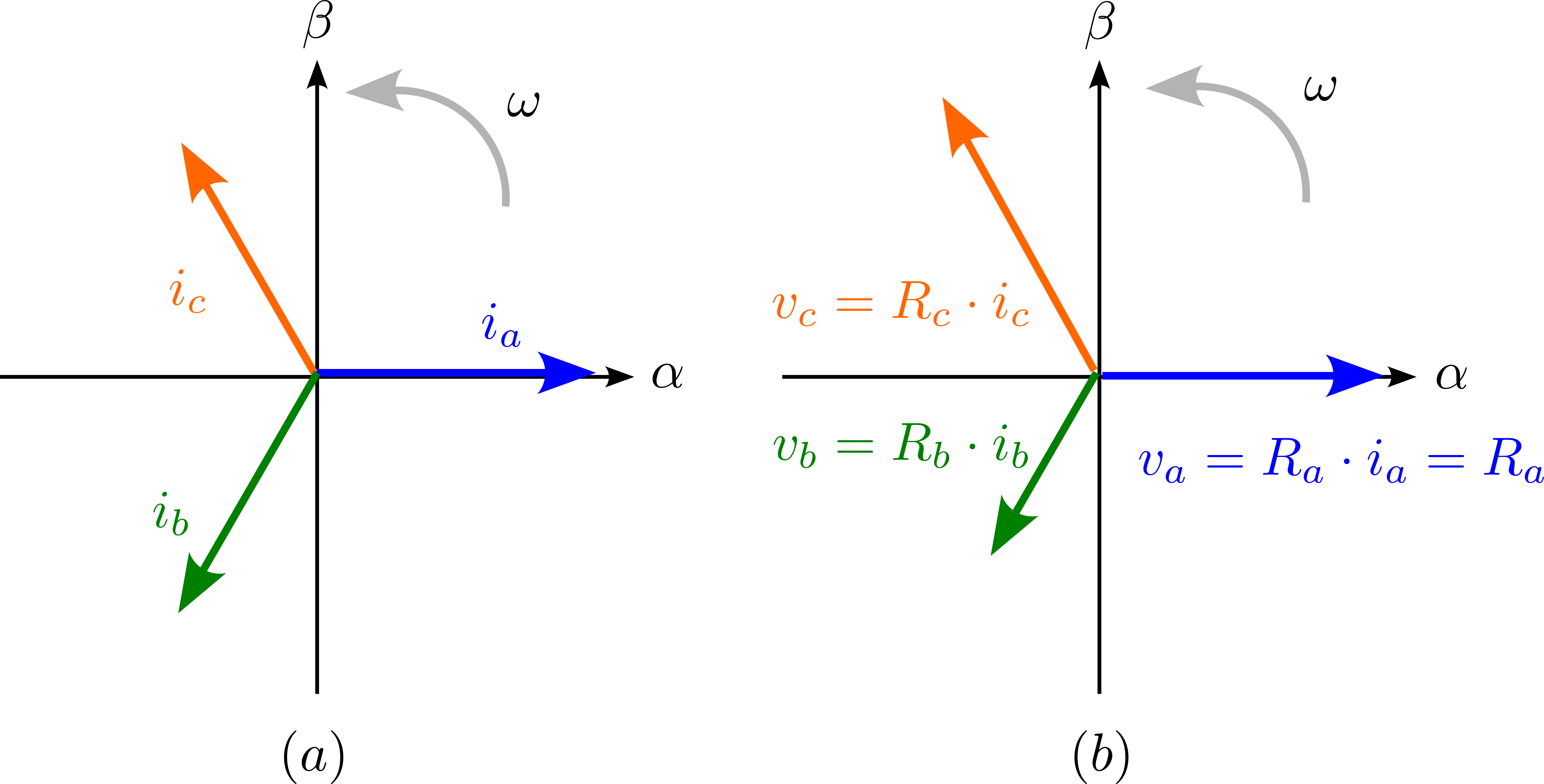}}
\caption{Phasor diagram of the three-phase signals for the calculation of the first column of the sequence-domain equivalent impedance. (a) Phasor diagram of a unit-value three-phase current with only d component of the positive sequence, (b) Phasor diagram of the resulting unbalanced three-phase voltage.}
\label{fasoresiv}
\end{figure}

In order to illustrate the process, the calculation using the sequence $i_d^+$ will be used. This case is the easiest to visualize, since injecting current in the $i_d^+$ sequence means injecting a perfectly balanced symmetrical current with a positive rotation sign. Figure \ref{phasores_R_Id} shows the phasor representations of the currents and voltages of the three-phase system on the orthogonal stationary $\alpha$ and $\beta$ axes. The sign of rotation of the three-phase system is also represented. It can be observed that the current phasors have the same magnitude, but the voltage phasors have different magnitudes, due to the different impedance values.

Next, the Lyon transformation (\ref{Lyon}) must be used to translate the system into positive, $v_a^+$, and negative sequence, $v_a^-$. It is not necessary to calculate the projections of phases $b$ and $c$, since they can be obtained directly by offsetting 120º from phase a of each sequence, taking into account that the positive and negative sequences are symmetrical by definition. To obtain the zero sequence, it is only needed to calculate the vector sum of the phases $a$, $b$, and $c$ of the original sequence. Notice that the zero sequence is contained in a vector, $\gamma$, that is perpendicular to the $\alpha \beta$ plane and an artificial vector in quadrature, $q\gamma$, must be generated to be able to represent the zero sequence in a plane and derive its $d$ and $q$ components. 

Defining a $dq$ framework that initially coincides with the $\alpha \beta$ axes, the representation of the different sequences in the $dq$ reference frame will be defined by finding the projection of the phase $a$ of each sequence onto the stationary $\alpha$ and $\beta$ axes.

\begin{figure}[htbp]
\centerline{\includegraphics[width=88mm]{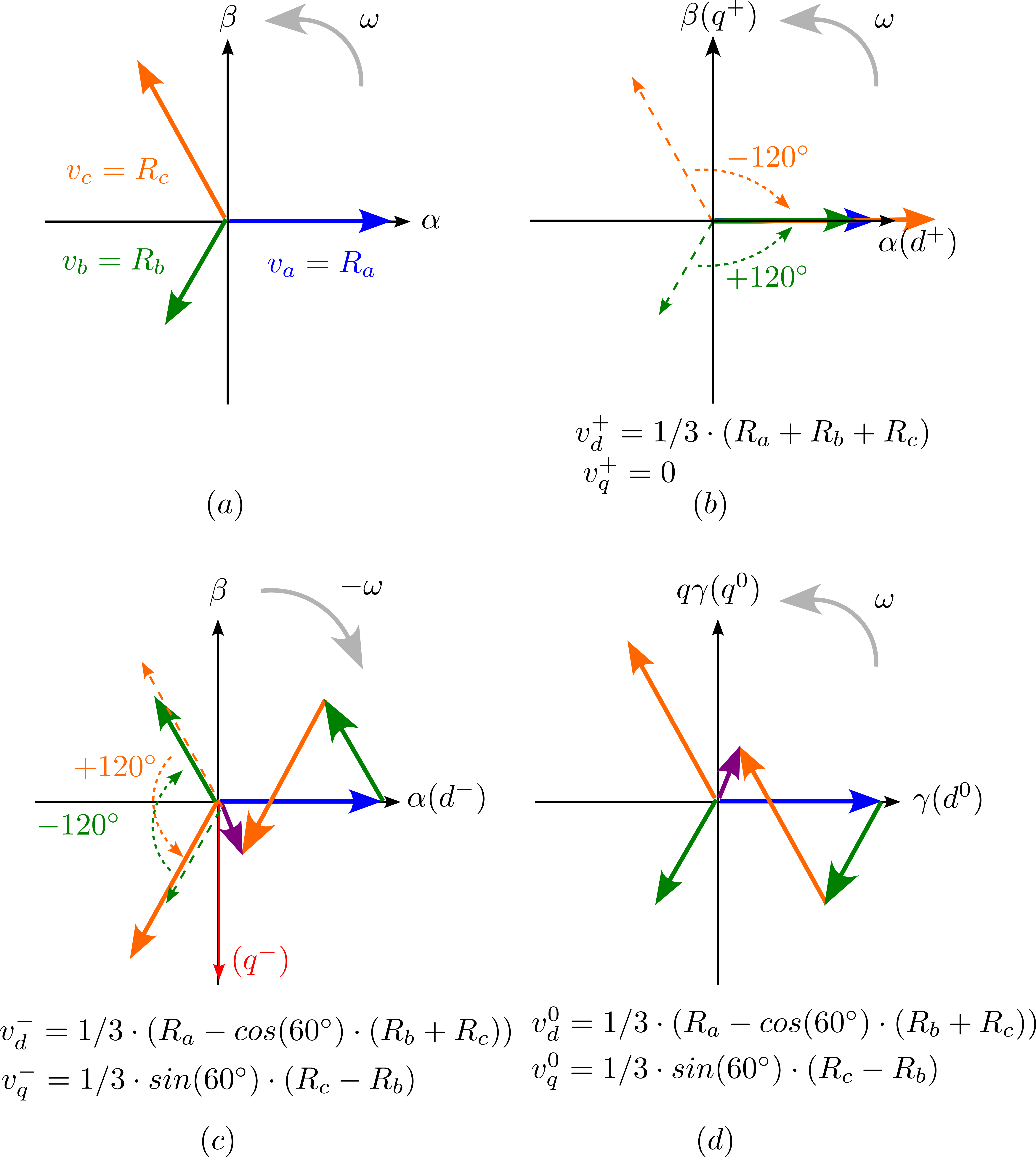}}
\caption{Derivation of the $d$ and $q$ components of the sequences of the voltage. (a) Phasor diagram of the unbalanced three-phase voltage when a unit-value current with only d component of the positive sequence is applied to the unbalanced resistive impedance, (b) Derivation of the $d$ and $q$ components of the positive sequence of the voltage, (c) Derivation of the $d$ and $q$ components of the negative sequence of the voltage, (d) Derivation of the $d$ and $q$ components of the zero sequence of the voltage.}
\label{fasoresidpos}
\end{figure}

In Fig.~\ref{fasoresidpos} it is illustrated how the projections of the sequences in the synchronous reference frame were calculated graphically. In $(a)$ the original unbalanced voltage phasors are represented. In $(b)$ the result of applying the Lyon transformation and obtaining the positive sequence is shown, whereas in $(c)$ and $(d)$ the negative and the zero transformation and sequences are represented, respectively. In these representations, the vectors shown with solid lines represent the situation of the vectors after applying the rotations determined by the corresponding row of the matrix $[T]$ (\ref{Lyon}). A dashed line is used to depict the original position of the voltage phasor, in order to facilitate the visualization of the rotation applied. 

After the rotation, the sum vector must be performed. In those figures in which the vectorial sum of the rotated phasors is not immediate, this sum is represented graphically, with the resulting sum vector being represented in purple. To calculate the projections in the reference system $dq$, the projection of the resulting vector (purple color) on the axes is taken. For example, $v_d^+$ and $v_q^+$ would be the projections on the d+ and q+ axes, respectively. The location of the axes of the reference frame $dq$ in each sequence is indicated on the axis that corresponds in each case. It is worth noting that in the case of the $q^-$ axis, this is on the negative side of the $\beta$-axis, being highlighted in red in (c). This is due to the negative sense of rotation of the negative sequence, opposite to that of the positive sequence $\omega$t.

The process of representation and graphical calculation stays the same for the rest of the cases. As it can be seen, it is possible to show that a three-phase current that is balanced, and therefore without a zero sequence component, can be applied to an unbalanced resistive impedance to generate an unbalanced three-phase voltage that has a zero sequence component.

\begin{figure*}[htbp]
\centerline{\includegraphics{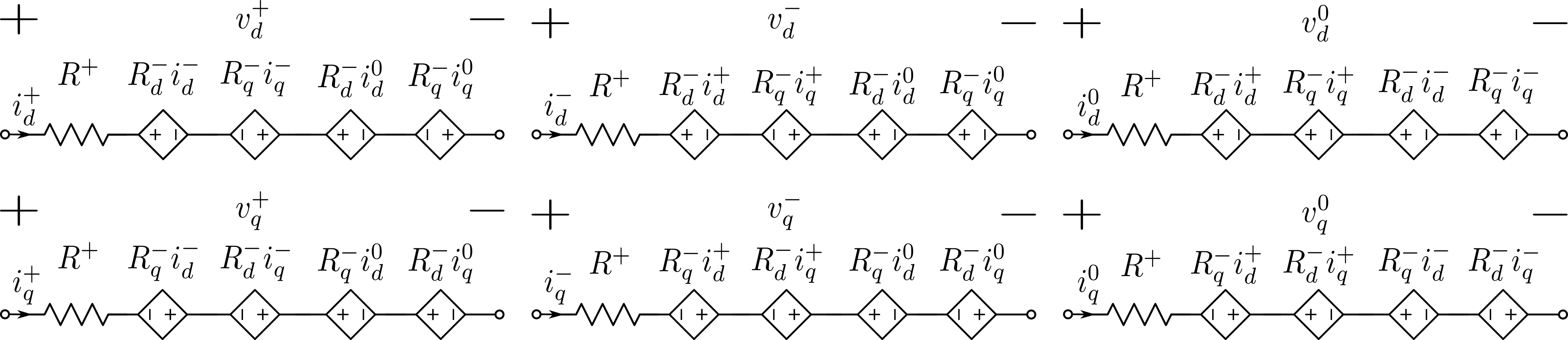}}
\caption{Sequence-domain equivalent circuit of an unbalanced three-phase resistive impedance.}
\label{eq_R}
\end{figure*} 

Applying these calculations for all the sequences, the equations for the equivalent circuit in the $dq$ reference frame can be obtained. For a purely resistive unbalanced three-phase impedance, with values $R_a$, $R_b$, and $R_c$, this circuit can be modeled by a single impedance matrix $[R^{+-0}_{dq}]$:

\begin{equation} \label{R_matrix}
[R^{+-0}_{dq}] =
\left[ \begin{array}{cccccc}
R^+ & 0 & R_d^- & -R_q^- & R_d^- & -R_q^-\\
0 & R^+ & -R_q^- & -R_d^- & R_q^- & R_d^-\\
R_d^- & -R_q^- & R^+ & 0 & R_d^- & R_q^-\\
-R_q^- & -R_d^- & 0 & R^+ & R_q^- & -R_d^-\\
R_d^- & R_q^- & R_d^- & R_q^- & R^+ & 0\\
-R_q^- & R_d^- & R_q^- & -R_d^- & 0 & R^+
\end{array} \right]
\end{equation}

\noindent where 

\begin{equation} \label{R_+}
R^+ = \dfrac{1}{3}\left(R_a+R_b+R_c\right)
\end{equation}
\begin{equation} \label{R_d}
R_d^- = \dfrac{1}{3}\left(R_a-\dfrac{R_b+R_c}{2}\right)
\end{equation}
\begin{equation} \label{R_q}
R_q^- = \dfrac{\sqrt{3}}{6}\left(R_b-R_c\right)
\end{equation}

In Fig.\ref{eq_R} the sequence-domain equivalent circuit of an unbalanced three-phase resistive impedance is depicted.

\subsection{Derivation of the Inductive Equivalent Impedance}

In the case of an inductive three-phase impedance, the circuit equation in the $abc$ reference frame can be expressed by:

\begin{equation} \label{Labc}
    \left[ \begin{array}{c}
v_a \\ \\ 
v_b \\ \\
v_c 
\end{array} \right]= 
\left[ \begin{array}{ccc}
L_a & 0 & 0 \\ \\
0 & L_b & 0 \\ \\ 
0 & 0 & L_c
\end{array} \right]
\left[ \begin{array}{c}
\dfrac{di_a}{dt} \\ \\
\dfrac{di_b}{dt} \\ \\
\dfrac{di_c}{dt} 
\end{array} \right] = \left[ L_{abc} \right] \cdot \dfrac{di_{abc}}{dt}
\end{equation}

In the case of a balanced impedance, the matrix $[L_{abc}]$ will be diagonal and $L_a=L_b=L_c=L$. When applying the Clarke transformation to the inductive impedance, the following equation is obtained:

\begin{equation} \label{L_en_ab}
     v_{\alpha \beta} = \left[ T_{c} \right] \left[ L_{abc}  \right] \left[ T^{-1}_{c} \right] \cdot \dfrac{d}{dt} ( i_{\alpha \beta}) = \left[ L_{\alpha \beta} \right]\cdot \dfrac{d}{dt} (i_{\alpha \beta})
\end{equation}

Then, applying the Park transformation leads to:

\begin{equation}
    v_{dq} \cdot e^{j\omega t} = \left[ L_{dq} \right] \cdot \dfrac{d}{dt}\cdot (i_{dq} \cdot e^{j\omega t})
\end{equation}

When operating the derivative term, the following is obtained:

\begin{equation}
    v_{dq} \cdot e^{j\omega t} = \left[ L_{dq} \right] \cdot \dfrac{di_{dq} }{dt}\cdot  e^{j\omega t} + \left[ L_{dq} \right] \cdot ( j\omega) \cdot i_{dq} \cdot  e^{j\omega t}
\end{equation}
\noindent where $j\omega$ can be represented in matrix form as:

\begin{equation}
    j\omega = \left[ \begin{array}{cc}
    0 & \omega  \\
    -\omega & 0   
     \end{array} \right]
\end{equation}

Finally, the phasor expression is:

\begin{equation}
    v_{dq} = \left[ L_{dq} \right] \cdot \dfrac{di_{dq} }{dt} + \left[ L_{dq} \right] \cdot \left[ \begin{array}{cc}
    0 & \omega  \\
    -\omega & 0   
     \end{array} \right] \cdot  i_{dq} 
\end{equation}

\noindent being:

\begin{equation}
   \left[ L_{dq} \right] =  \left[ L_{\alpha \beta} \right] = L \cdot I
\end{equation}


For an unbalanced impedance, the symmetrical components must be obtained, extending the impedance matrix to account for its contribution to each sequence:

\begin{equation} \label{L_seq}
[v^{+-0}_{abc}] = [L^{+-0}_{abc}] \cdot \dfrac{di^{+-0}_{abc}}{dt}
\end{equation}

Then applying the Clarke transformation:

\begin{equation} \label{L_clarke}
[v^{+-0}_{\alpha \beta}] = \left[ L^{+-0}_{\alpha \beta}  \right] \cdot \dfrac{di^{+-0}_{\alpha \beta}}{dt}
\end{equation}

Finally, the Park transformation is applied. Note that for the positive and zero sequences, $T_p(\omega)$ is used (see (\ref{Park2})), while in the negative sequence $T_p(-\omega)$ is used since the negative reference frame can be considered as rotating in the opposite direction.

\begin{figure*}[htbp]
\centering
{\includegraphics[width=120mm]{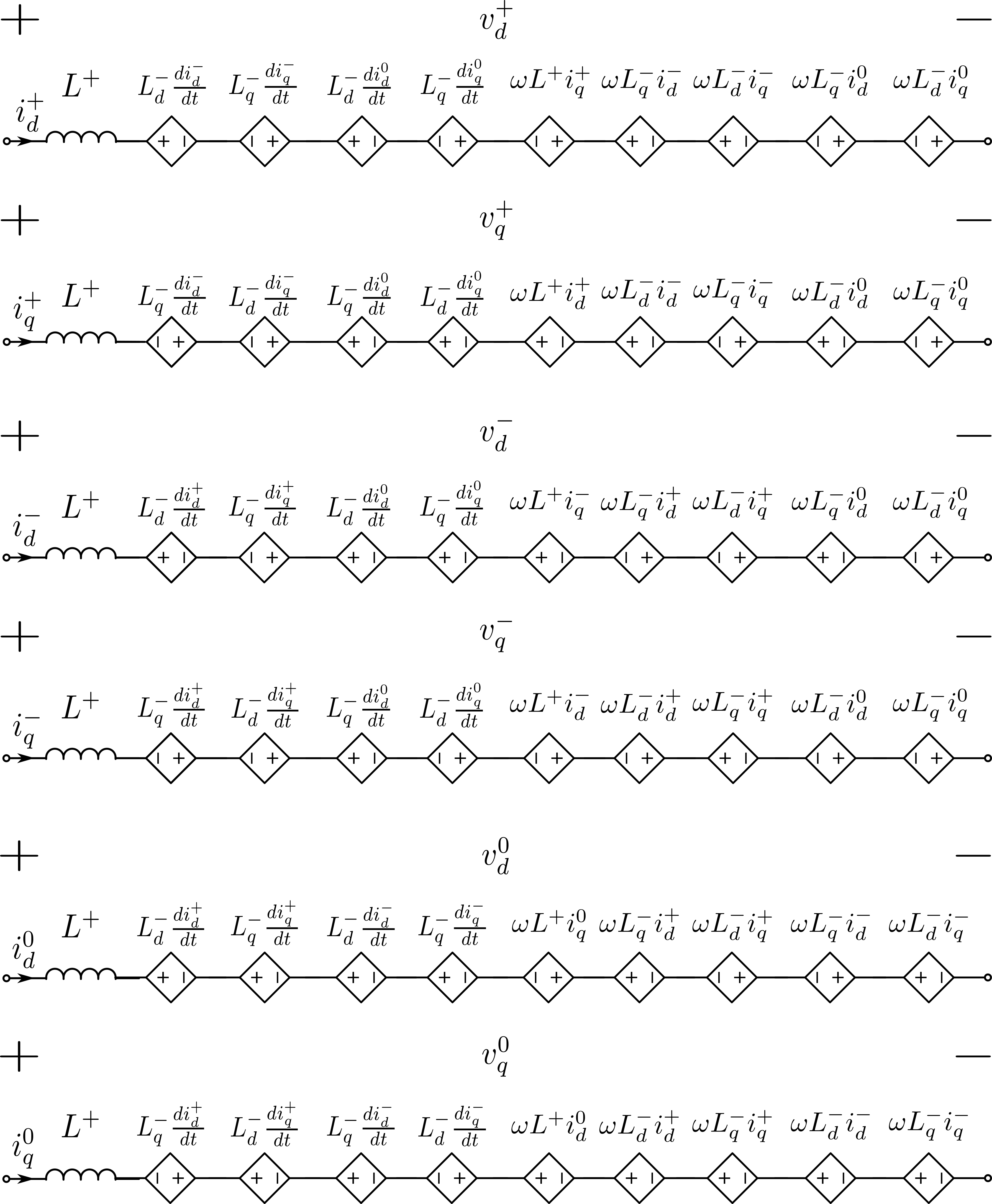}}
\caption{Sequence-domain equivalent circuit of an unbalanced three-phase inductive impedance.}
\label{eq_L}
\end{figure*} 

The phasor equation for the equivalent circuit in the sequence domain is given by:

\begin{equation} \label{L_equivalent}
   v_{dq}^{+-0}=[L^{+-0}_{dq}] \cdot \dfrac{di_{dq}^{+-0}}{dt}+ [L^{+-0}_{dq}]\cdot[\Omega] \cdot i_{dq}^{+-0}
\end{equation}

\noindent where 
\begin{equation} \label{L_matrix}
[L^{+-0}_{dq}] =
\left[ \begin{array}{cccccc}
L^+ & 0 & L_d^- & -L_q^- & L_d^- & -L_q^-\\
0 & L^+ & -L_q^- & -L_d^- & L_q^- & L_d^-\\
L_d^- & -L_q^- & L^+ & 0 & L_d^- & L_q^-\\
-L_q^- & -L_d^- & 0 & L^+ & L_q^- & -L_d^-\\
L_d^- & L_q^- & L_d^- & L_q^- & L^+ & 0\\
-L_q^- & L_d^- & L_q^- & -L_d^- & 0 & L^+
\end{array} \right]
\end{equation}

\begin{equation} \label{L_+}
L^+ = \dfrac{1}{3}\left(L_a+L_b+L_c\right)
\end{equation}
\begin{equation} \label{L_d}
L_d^- = \dfrac{1}{3}\left(L_a-\dfrac{L_b+L_c}{2}\right)
\end{equation}
\begin{equation} \label{L_q}
L_q^- = \dfrac{\sqrt{3}}{6}\left(L_b-L_c\right)
\end{equation}

\noindent and
\begin{equation} \label{omega}
[\Omega] =
\left[ \begin{array}{cccccc}
0 & \omega & 0 & 0 & 0 & 0\\
-\omega & 0 & 0 & 0 & 0 & 0\\
0 & 0 & 0 & -\omega & 0 & 0\\
0 & 0 & \omega & 0 & 0 & 0\\
0 & 0 & 0 & 0 & 0 & \omega\\
0 & 0 & 0 & 0 & -\omega & 0\\
\end{array} \right]
\end{equation}
The extended inductance matrix $[L^{+-0}_{dq}]$ is calculated using the same process from the resistive impedance scenario. Therefore, its terms are completely analogous to those in $[R^{+-0}_{dq}]$.

The sequence-domain equivalent circuit of an unbalanced three-phase inductive impedance is depicted in Fig. \ref{eq_L}.

\subsection{Derivation of the Capacitive Equivalent Impedance}

\begin{figure*}[htbp]
\centering
{\includegraphics[width=180mm]{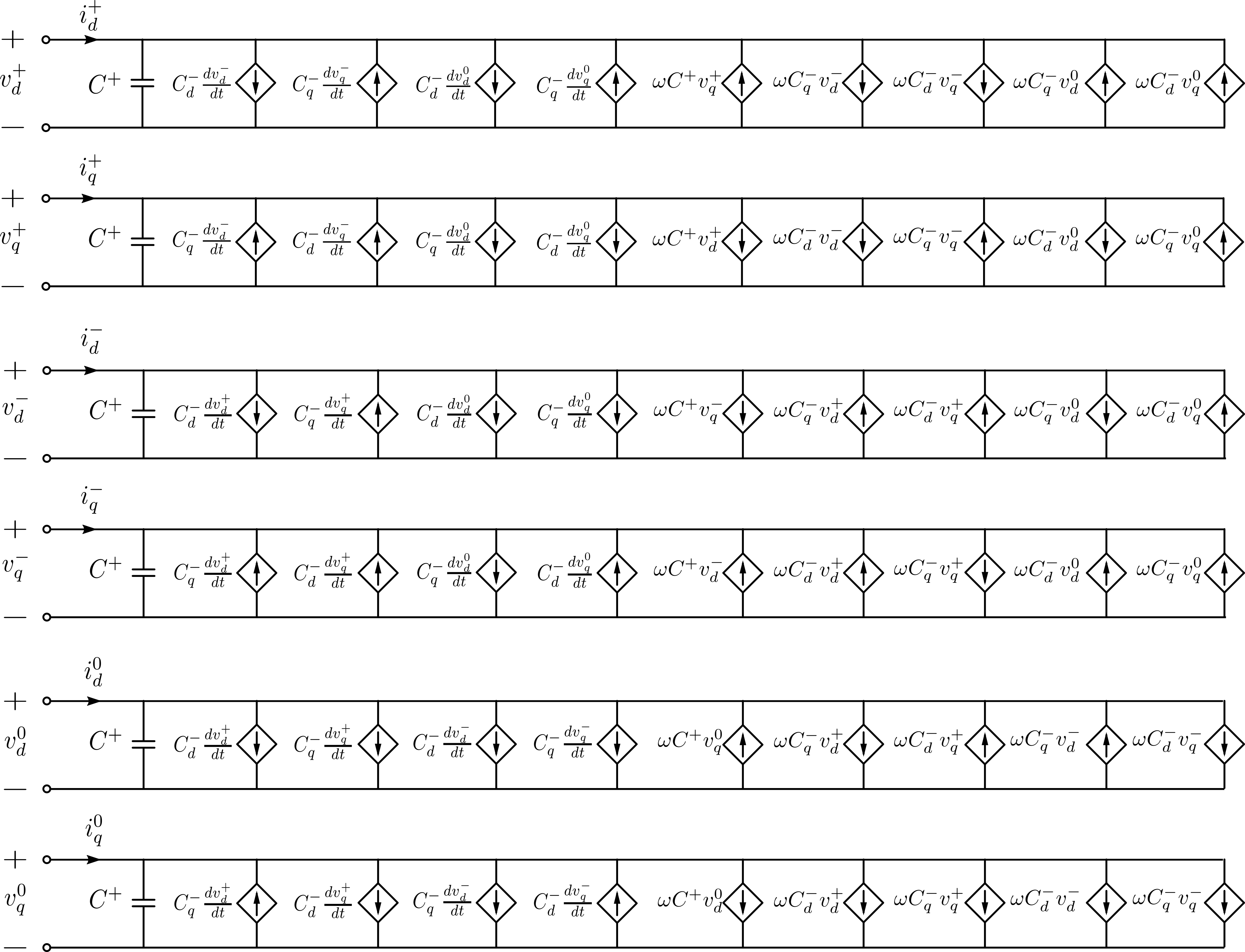}}
\caption{Sequence-domain equivalent circuit of an unbalanced three-phase capacitive impedance.}
\label{eq_C}
\end{figure*}

The case of a capacitive three-phase impedance is analogous to the inductive impedance. In this case, the relationship must be established between the current and the voltage derivative as given by the equation of the capacitive impedance:

\begin{equation} \label{Cabc}
    \left[ \begin{array}{c}
i_a \\
i_b \\
i_c 
\end{array} \right]= 
\left[ \begin{array}{ccc}
C_a & 0 & 0 \\ \\
0 & C_b & 0 \\ \\
0 & 0 & C_c
\end{array} \right]
\left[ \begin{array}{c}
\dfrac{dv_a}{dt} \\ \\
\dfrac{dv_b}{dt} \\ \\
\dfrac{dv_c}{dt}
\end{array} \right] = \left[ C_{abc} \right] \cdot \dfrac{dv_{abc}}{dt}
\end{equation}

Following the established derivation method, the phasor equation for the equivalent circuit in the sequence domain can be obtained:

\begin{equation} \label{C_equivalent}
   i_{dq}^{+-0}=[C^{+-0}_{dq}] \cdot \dfrac{dv_{dq}^{+-0}}{dt}+ [C^{+-0}_{dq}]\cdot[\Omega] \cdot v_{dq}^{+-0}
\end{equation}

\noindent where 
\begin{equation} \label{C_matrix}
[C^{+-0}_{dq}] =
\left[ \begin{array}{cccccc}
C^+ & 0 & C_d^- & -C_q^- & C_d^- & -C_q^-\\
0 & C^+ & -C_q^- & -C_d^- & C_q^- & C_d^-\\
C_d^- & -C_q^- & C^+ & 0 & C_d^- & C_q^-\\
-C_q^- & -C_d^- & 0 & C^+ & C_q^- & -C_d^-\\
C_d^- & C_q^- & C_d^- & C_q^- & C^+ & 0\\
-C_q^- & C_d^- & C_q^- & -C_d^- & 0 & C^+
\end{array} \right]
\end{equation}

\begin{equation} \label{C_+}
C^+ = \dfrac{1}{3}\left(C_a+C_b+C_c\right)
\end{equation}
\begin{equation} \label{C_d}
C_d^- = \dfrac{1}{3}\left(C_a-\dfrac{C_b+C_c}{2}\right)
\end{equation}
\begin{equation} \label{C_q}
C_q^- = \dfrac{\sqrt{3}}{6}\left(C_b-C_c\right)
\end{equation}

The rotation matrix $[\Omega]$ is as expressed in (\ref{omega}).

In Fig. \ref{eq_C} the sequence-domain equivalent circuit of an unbalanced three-phase capacitive impedance is depicted. 

\section{Experimental Validation}

\begin{figure*}[htbp]
\centerline{\includegraphics[width=130mm]{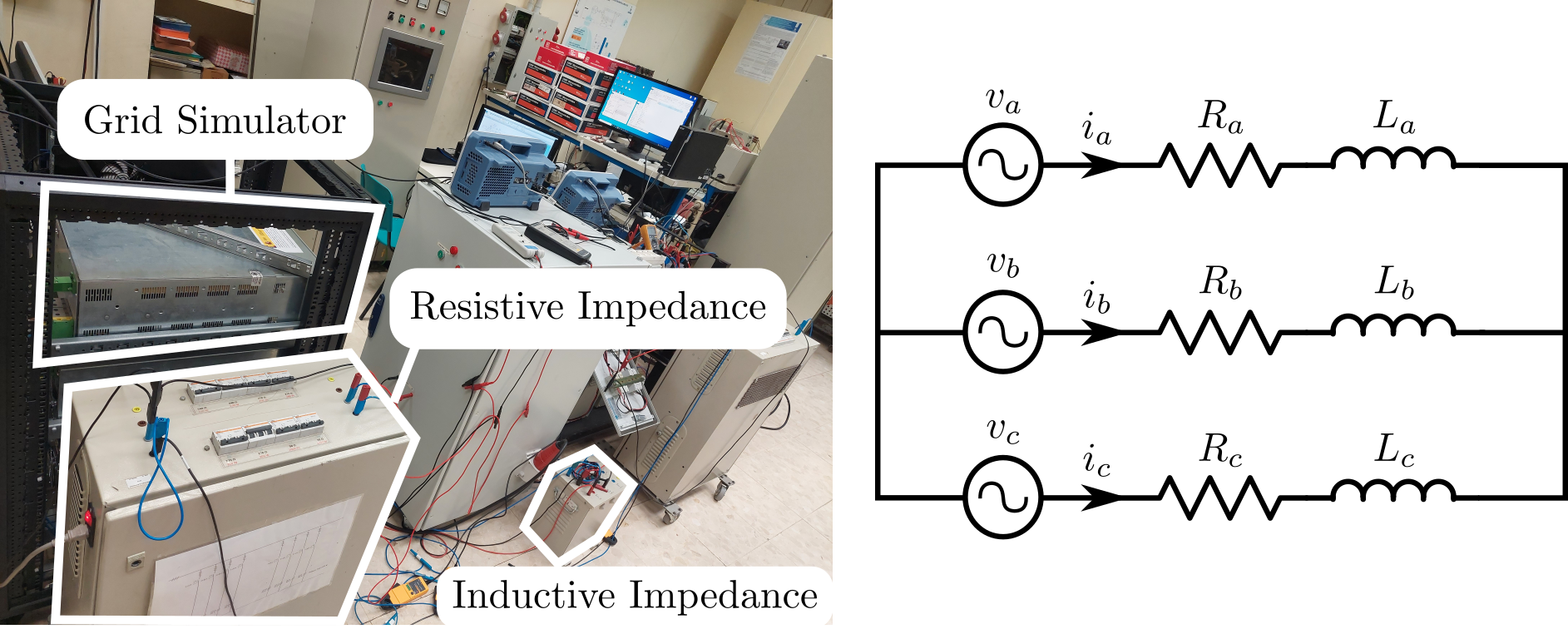}}
\caption{ Experimental setup. Left: picture of the setup, right: electrical circuit diagram.}
\label{setup}
\end{figure*}

The proposed equivalent model has been validated by comparing simulations with theoretical parameters along with experimental results. The tests were carried out by generating an asymmetric three-phase voltage, by means of a four-quadrant grid simulator (REGATRON TC.ACS), and applying it to  a three-phase unbalanced RL impedance (see Fig. \ref{setup}). The impedance values used in this particular test are $R_a=50$~$\Omega$, $R_b=270$~$\Omega$, $R_c=270$~$\Omega$, $L_a=32$~$mH$, $L_b=32$~$mH$, and $L_c=16$~$mH$.

\begin{figure*}[htbp]
\centerline{\includegraphics{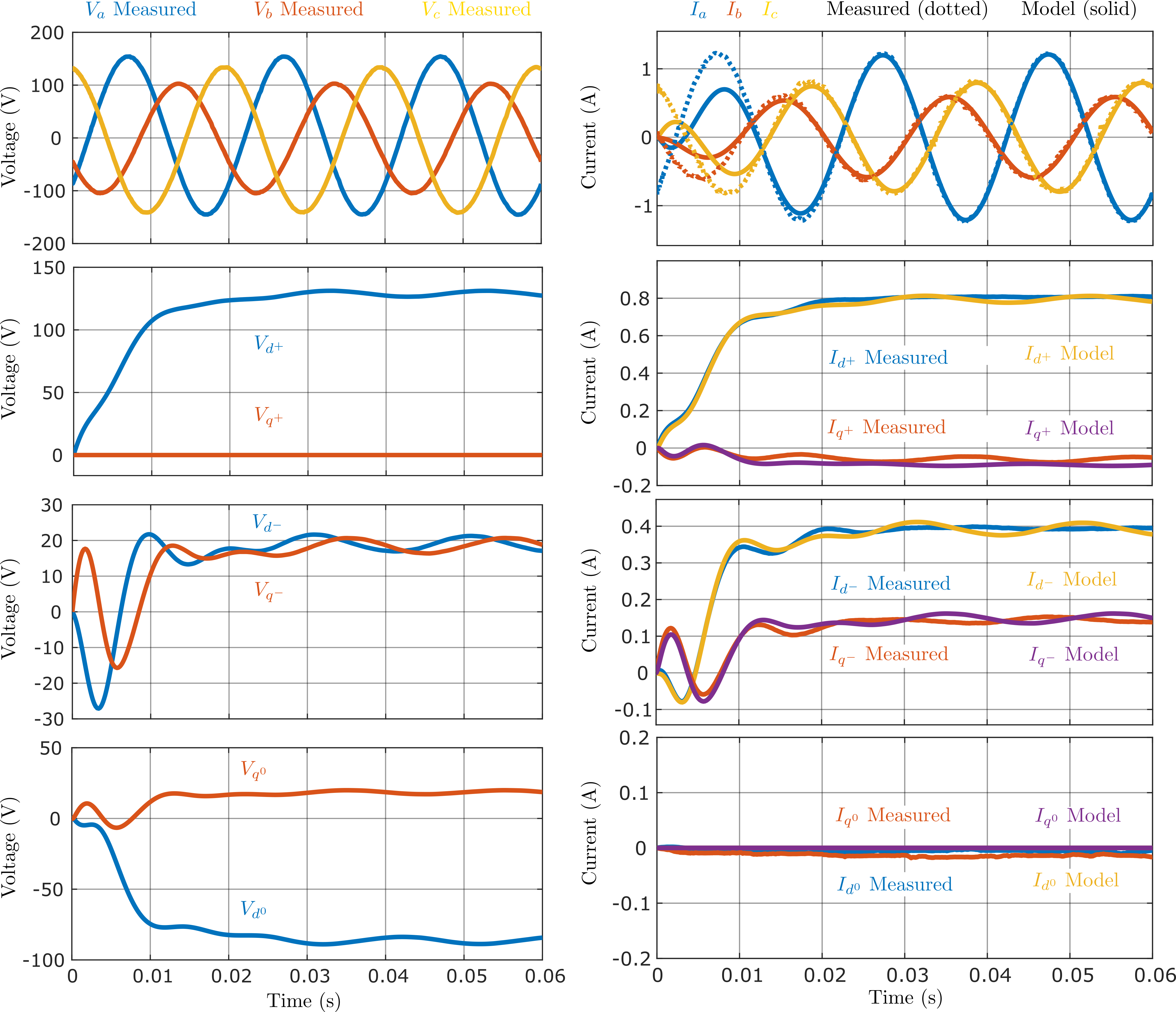}}
\caption{Comparison between measured and simulated signals. The $abc$ and $dq$ sequence components of the signals are shown. Left: imposed voltages, right: measured and simulated resulting currents.}
\label{comparison}
\end{figure*}

Fig. \ref{comparison} shows the applied voltages and the comparison between the measured current against the calculated current using the proposed equivalent circuit model. On the left side of the figure, the applied voltage is represented, the first figure shows the three-phase voltage, the second the $d$ and $q$ components of the positive sequence, the third the $d$ and $q$ components of the negative sequence, and the fourth the $d$ and $q$ components of the zero sequence. Notice that the applied voltage is balanced, but the zero sequence is the voltage that appears between neutral points due to the unbalanced impedance, which can be predicted by the proposed model. 

On the right side of the figure, the measured current is compared with the predicted current of the proposed model, the first figure shows the three-phase currents, the second the $d$ and $q$ components of the positive sequence, the third the $d$ and $q$ components of the negative sequence, and the fourth the $d$ and $q$ components of the zero sequence. It can be seen that the equivalent model is able to reproduce the $dq$ sequence-domain components of the resulting current seamlessly.

Notice that to obtain the symmetric components from the $abc$ signals, a Second Order Generalized Intergrator Quadrature Signal Generator (SOGI-QSG) has been implemented \cite{Teodorescu_Book_2010}. This element has a filter effect in the derivation of the sequences and that is the reason for the transient responses and the initial mismatch in the current prediction.

\section{Conclusions}

This paper proposes an equivalent $dq$ sequence-domain model of unbalanced three-phase passive impedances to improve the design of controllers for three-phase power converters. The main advantage of this framework, from the control point of view, is that all the components are constant in the steady state, even if both the three-phase impedance and voltage or current are asymmetrical and unbalanced. Furthermore, the proposed equivalent model explicitly shows the interaction among the different sequences, including the zero sequence, due to the fact that the three-phase impedance is unbalanced, which was not properly considered in previous studies. The proposed equivalent model has been validated with experimental results.

\bibliographystyle{IEEEtran}
\bibliography{biblio}

@ARTICLE{Control_stationary,
  author={Wodyk, Sebastian and Iwanski, Grzegorz},
  journal={IEEE Transactions on Industry Applications}, 
  title={Control of Three-Phase Power Electronic Converter With Power Controllers in Stationary Frame}, 
  year={2020},
  volume={56},
  number={5},
  pages={5257-5268},
  doi={10.1109/TIA.2020.3009311}}

@ARTICLE{grid_voltage_estimation_stationary,
  author={Chowdhury, Vikram Roy},
  journal={IEEE Transactions on Energy Conversion}, 
  title={Internal Model Based Grid Voltage Estimation and Control of a Three-Phase Grid Connected Inverter for $PV$ Application}, 
  year={2021},
  volume={36},
  number={4},
  pages={3568-3577},
  doi={10.1109/TEC.2021.3079908}}

@ARTICLE{resonant_control_stationary,
  author={Hu, Mingjin and Hua, Wei and Ma, Guangtong and Xu, Shuai and Zeng, Weitong},
  journal={IEEE Transactions on Industrial Electronics}, 
  title={Improved Current Dynamics of Proportional-Integral-Resonant Controller for a Dual Three-Phase FSPM Machine}, 
  year={2021},
  volume={68},
  number={12},
  pages={11719-11730},
  doi={10.1109/TIE.2020.3047009}}

@ARTICLE{Input_voltage_distortion_dq,
  author={Trinh, Quoc Nam and Wang, Peng and Choo, Fook Hoong},
  journal={IEEE Journal of Emerging and Selected Topics in Power Electronics}, 
  title={An Improved Control Strategy of Three-Phase PWM Rectifiers Under Input Voltage Distortions and DC-Offset Measurement Errors}, 
  year={2017},
  volume={5},
  number={3},
  pages={1164-1176},
  doi={10.1109/JESTPE.2017.2670229}}

@ARTICLE{virtual_time_constant_dq,
  author={Kukrer, Osman and Bayhan, Sertac and Komurcugil, Hasan},
  journal={IEEE Transactions on Industrial Electronics}, 
  title={Model-Based Current Control Strategy With Virtual Time Constant for Improved Dynamic Response of Three-Phase Grid-Connected VSI}, 
  year={2019},
  volume={66},
  number={6},
  pages={4156-4165},
  doi={10.1109/TIE.2018.2863195}}

@INPROCEEDINGS{decentralized_control_dq,
  author={Ghosh, Subhrasankha and Chattopadhyay, Souvik and Samanta, Sayan},
  booktitle={2019 IEEE Applied Power Electronics Conference and Exposition (APEC)}, 
  title={A Synchronous Reference Frame based Decentralized Control Architecture for Inverters Connected to an Autonomous Microgrid}, 
  year={2019},
  volume={},
  number={},
  pages={540-547},
  doi={10.1109/APEC.2019.8721904}}

@ARTICLE{impedance_modeling_sequence,
  author={Cespedes, Mauricio and Sun, Jian},
  journal={IEEE Transactions on Power Electronics}, 
  title={Impedance Modeling and Analysis of Grid-Connected Voltage-Source Converters}, 
  year={2014},
  volume={29},
  number={3},
  pages={1254-1261},
  doi={10.1109/TPEL.2013.2262473}}

@INPROCEEDINGS{Model_predictive_sequence,
  author={Li, Xiaoyan and Zhang, Chenghui and Chen, Alian and Xing, Xiangyang and Zhang, Guangxian},
  booktitle={2018 IEEE Applied Power Electronics Conference and Exposition (APEC)}, 
  title={Model predictive direct current control strategy for three-level T-type rectifier under unbalanced grid voltage conditions}, 
  year={2018},
  volume={},
  number={},
  pages={1514-1519},
  doi={10.1109/APEC.2018.8341217}}

@ARTICLE{Dual_current_control_sequence,
  author={Azzouz, Maher A. and Hooshyar, Ali},
  journal={IEEE Transactions on Smart Grid}, 
  title={Dual Current Control of Inverter-Interfaced Renewable Energy Sources for Precise Phase Selection}, 
  year={2019},
  volume={10},
  number={5},
  pages={5092-5102},
  doi={10.1109/TSG.2018.2875422}}

@article{unbalanced_impedance_original,
author = {Li, Zhen and Wong, Siu-Chung and Tse, Chi K. and Liu, Xiangdong},
title = {Modeling of unbalanced three-phase driving-point impedance with application to control of grid-connected power converters},
journal = {International Journal of Circuit Theory and Applications},
volume = {44},
number = {4},
pages = {851-873},
doi = {https://doi.org/10.1002/cta.2110},
url = {https://onlinelibrary.wiley.com/doi/abs/10.1002/cta.2110},
eprint = {https://onlinelibrary.wiley.com/doi/pdf/10.1002/cta.2110},
year = {2016}
}

@INPROCEEDINGS{adaptive_control_unbalanced_impedance,
  author={Lu, Zelun and Li, Wenxuan and Li, Zhen and Chen, Xi and Lu, Herbert H. C. and Dong, Ning and Liu, Xiangdong},
  booktitle={2017 IEEE International Symposium on Circuits and Systems (ISCAS)}, 
  title={Adaptive droop control with self-adjusted virtual impedance for three-phase inverter under unbalanced conditions}, 
  year={2017},
  volume={},
  number={},
  pages={1-4},
  doi={10.1109/ISCAS.2017.8050928}}

@book{Teodorescu_Book_2010,
author = {Teodorescu, Remus and Liserre, Marco and Rodr{\'{i}}guez, Pedro},
booktitle = {Grid Converters for Photovoltaic and Wind Power Systems},
doi = {10.1002/9780470667057},
isbn = {9780470057513},
publisher = {John Wiley and Sons},
title = {{Grid Converters for Photovoltaic and Wind Power Systems}},
year = {2010}
}

\end{document}